\documentclass[runningheads]{llncs}
\usepackage{graphicx}
\usepackage{subfig}
\usepackage{float}
\usepackage{amsmath}
\usepackage{longtable}
\usepackage{hyperref}
\usepackage{listings}

\begin{document}
%
\title{Unlocking Insights: Semantic Search in Jupyter Notebooks\thanks{Supported by Microsoft Research.}}
\titlerunning{Unlocking Insights: Semantic Search in Jupyter Notebooks}

\author{Lan Li\inst{1}\orcidID{0000-0003-4499-4126} \and
Jinpeng Lv\inst{2}}

\authorrunning{Lan Li et al.}
%
\institute{Center for Informatics Research in Science \& Scholarship \\
School of Information Sciences\\University of Illinois at Urbana-Champaign, USA\\
\email{lanl2@illinois.edu}\\
\and
Research System Team, Microsoft, Redmond, USA\\
\email{jinpeng.lv@microsoft.com}}
\maketitle              

\begin{abstract}

Semantic search, a process aimed at delivering highly relevant search results by comprehending the searcher's intent and the contextual meaning of terms within a searchable dataspace, plays a pivotal role in information retrieval. In this paper, we investigate the application of large language models to enhance semantic search capabilities, specifically tailored for the domain of Jupyter Notebooks. Our objective is to retrieve generated outputs, such as figures or tables, associated functions and methods, and other pertinent information.

We demonstrate a semantic search framework that achieves a comprehensive semantic understanding of the entire notebook's contents, enabling it to effectively handle various types of user queries. Key components of this framework include:

1). A data preprocessor is designed to handle diverse types of cells within Jupyter Notebooks, encompassing both markdown and code cells. \\
2). An innovative methodology is devised to address token size limitations that arise with code-type cells. We implement a finer-grained approach to data input, transitioning from the cell level to the function level, effectively resolving these issues.


\end{abstract}

\section{Introduction}

In contrast to traditional keyword-based searches, which seek literal matches or variations of query terms without full comprehension of their contextual meaning, semantic search requires an understanding of the searcher's intent and the contextual meaning of terms within the searching space. Embedding generated by language models has been applied to various areas, such as finding relatedness among entities in Knowledge Graphs (KGs) \cite{morales}, generating KGs from semi-structured data, querying, and managing Knowledge Graph Engineering (KGE) \cite{meyer}, and comparing whether records from a pair of relational tables refer to the same entity or not \cite{fl:bl}. All these applications have demonstrated that embeddings can appropriately reflect the semantics for different types of data, ranging from unstructured, semi-structured, to structured datasets.

Jupyter Notebook empowers users to create narratives comprising live, re-executable code, mathematical equations, descriptive text, interactive dashboards, and various multimedia elements \cite{klu:will}. Therefore, a single Jupyter Notebook, usually loaded in JSON format, is composed of various types of information. With the widespread adoption of large language models (LLMs) across diverse applications, yielding remarkable results, we extend an invitation to LLMs to play a pivotal role in enhancing and empowering the semantic search process within the domain of Jupyter Notebooks. 

Based on our study on user requirements for searching Jupyter Notebooks within the team, users are interested in exploring data visualizations, analyzing plotted results, accessing tutorials for specific method usage, inspecting code implementations of ad-hoc functions, and reviewing general output information from the notebook contents. These requirements depend on a comprehensive and nuanced semantic grasp of Jupyter Notebook. In a nutshell, our assumption is that LLMs can comprehend and generate meaningful embeddings for Jupyter Notebooks, thereby supporting the query process in a robust manner.

In summary, this project is driven by several key goals, which include exploring the potential of LLMs and addressing challenges from the following perspectives:

\begin{itemize}
    \item Identifying Data Inputs for Jupyter Notebook content embeddings: pinpoint the essential data inputs necessary for generating embeddings from Jupyter Notebook content.

    \item Resolving token size limitation restricted by embedding model: propose a finer-grained granularity of data inputs (code-type cell) from cell level to function level.

    \item Studying code summaries by the latest GPT-4 model: study and confirm the preservation of information when transitioning from the original code to code summaries generated by the GPT-4 model.
    
    \item Evaluating the ability of addressing users' queries: accommodate various types of user queries comprehensively.
\end{itemize}

\section{Related Work: Transformers in Semantic Search and Code Summarization}

Semantic search is composed of two tasks: 1. understanding the \textit{semantics} of queries and documents beyond keywords; 2. finding top k answers from a document corpus given a query as \textit{search} result \cite{muen}. With our case, semantic search for Jupyter Notebooks, a JSON format data which is composed of both textual part (i.e., markdown) and code part, requires a good understanding of both types of contents. 

Transformers are widely applied to Natural Language Process tasks, also shed lights on semantic search area. Search applications like Google \cite{nayak} and Bing \cite{jeff:cass} employ transformers to provide semantically meaningful search results. However, it's important to note that these applications utilize limited BERT-like encoder-only transformers. In contrast, Muennighoff et al. \cite{muen} introduce a different approach by applying decoder-only transformers, referred to as SGPT, for semantic search and the generating meaningful sentence embeddings. The primary emphasis of the paper lies in evaluating SGPT's performance on the BEIR search benchmark, raising uncertainty about its effectiveness in addressing various other semantic search tasks \cite{koubaa:latif}.

GPT-based models hold substantial promise for application in diverse domains, encompassing education, medicine, history, mathematics, physics, and beyond. These models can streamline various tasks, such as generating summaries, answering questions, and offering personalized recommendations for users \cite{liu:ge}. In particular, The foundation models pretrained on massive data and finetuned to downstream tasks have been widely applied for code summarization, reducing the burden of software development and maintenance \cite{gu:gall}. 

Code summarization is the task of generating descriptions in natural language to document the implementation details, data inputs, and data outputs, which plays an important role in understanding the semantics of code snippets \cite{gu:gall} (See Table \ref{example:code-sum}). In contrast to the automatic code summarization models supported by CodeBERT and GPT-2, as discussed by \cite{gu:gall}, we leverage the state-of-the-art GPT-4 \footnote{https://platform.openai.com/docs/models/gpt-4} model for code summarization. GPT-4 model allows for a larger token size input, training on more extensive datasets, and the ability to generate more complex responses.

\begin{table}
\centering
\vspace{-2em}
\caption{One example code from the LeetCode website: the first solution from \href{https://leetcode.com/problems/two-sum/solutions/737092/sum-megapost-python3-solution-with-a-detailed-explanation/}{two-sum} and code summary generated by \emph{gpt-4-32k} model. Code summary generated by the Language Model (LLM) is well-structured and provides in-depth explanations of its internal logic and mechanisms.}\label{example:code-sum}
\begin{tabular}{|p{2.5cm}|p{9cm}|}
\hline
Code &  
\small
\begin{lstlisting}[language=Python]
class Solution:
   def twoSum(self, nums: List[int], target: 
       int) -> List[int]:
       seen = {}
       for i, value in enumerate(nums): #1
           remaining = target - nums[i] #2
           
           if remaining in seen: #3
               return [i, seen[remaining]]  #4
           else:
               seen[value] = i  #5

\end{lstlisting} \\
\hline

 Code Summary&  The given code \textbf{defines a class} Solution with a \textbf{method} twoSum. This method takes a list of integers (numbers) and an integer target as \textbf{input} and \textbf{returns} a list of two indices from numbers that add up to the target. It \textbf{uses a dictionary (seen)} to store the numbers and their indices encountered so far, and \textbf{calculates the remaining value} needed to \textbf{reach the target}. If this remaining value is found in the dictionary, it returns the corresponding indices. \\
\hline  
\end{tabular}
\end{table}

\section{Approach and Demonstration}

\begin{figure}[!ht]%
    \centering
    \subfloat[\centering This is the overall architecture, consisting of: 1). Embedding and Completion models are used to generate embeddings for cells of each Jupyter Notebook file; 2). Data insertion and database update: save the cell contents and corresponding embeddings vectors into a vector database named Weaviate \cite{weaviate}, and update it every fifteen minutes; 3). Search engine: similarity search and return top k related results. ]{{\includegraphics[width=5cm]{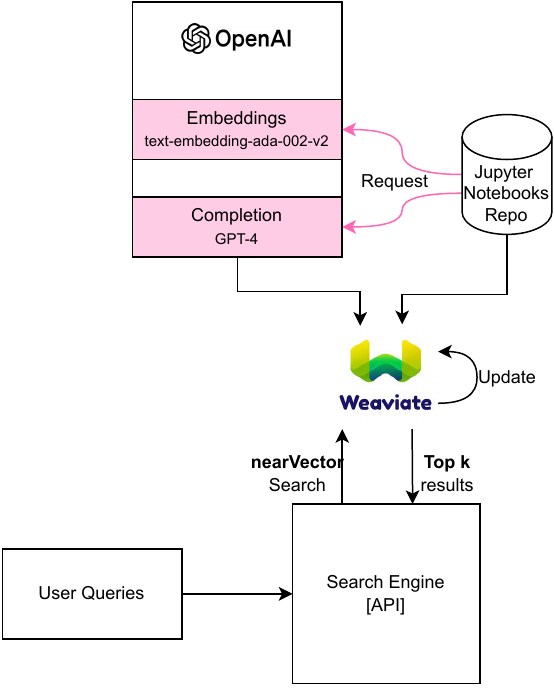} }}%
    \qquad
    \subfloat[\centering Our embedding generation pipeline consists of three stages: Document chunking: Dividing the Jupyter Notebook file into cells. Data preprocessing: Employing distinct preparation methods for markdown and code cells. Requesting GPT models to generate embeddings.]{{\includegraphics[width=5cm]{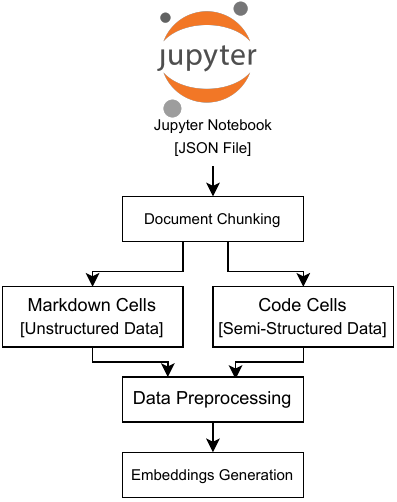} }}%
    \caption{Architecture of Semantic Search Framework (Left) and Pipeline of Embedding Generation (Right)}%
    \label{semantic-search}
\end{figure}

We develop a semantic search framework for Jupyter Notebooks (see the architecture on the left and the detailed pipeline of embedding generation process on the right of Figure \ref{semantic-search}). 

\paragraph{Data Preprocessing} 

The most time-consuming and intricate aspect of our project lies in the process of generating embeddings. To ensure the effectiveness of our semantic search system for Jupyter Notebooks, we began by conducting a comprehensive survey to gather user requirements regarding the search functionality. We decide to focus on two fundamental types of cells within Jupyter Notebooks accordingly: markdown cells, which typically contain descriptive overviews of the notebook's contents, and code cells, where implementation details are documented.

Once we segmented the Jupyter Notebook files into these cells, various data preprocessing methods are provided for each cell type. For markdown cells, our preparators would remove various elements such as hyperlinks, special characters (e.g, \{,\},:,``,",',!), and meaningless JSON field names. The well-documented and explained markdown cells greatly enhances the semantic value of the notebook's content. However, we also recognized the diversity in documentation styles among domain experts, data scientists, and engineers who contribute to the Jupyter Notebook repository. In some cases, notebooks did not contain markdown cells at all, but utilize separate files for documentation purposes or instructions.

\paragraph{Complete Embedding}
In addressing these challenging situations, we recognized the necessity of adapting our approach to generate embeddings exclusively from the code cells. The primary challenge we encounter is by the token size limitation issues imposed by our chosen embedding model, namely, \emph{text-embedding-ada-002-v2}, which allows for a maximum of 8191 tokens as input. There are several failed practical use cases list as follows: 1). Long manual input of data values within code cells. 2). Classes and functions are often consolidated into a single cell in case of tutorial notebooks. 3). Functions are of intricate and complex structures within individual cells.

To condense the code contents while retaining essential information, we switch from creating embeddings at the cell level to a finer-grained approach that focuses on individual functions. Firstly, we apply the Abstract Syntax Tree (AST), an Python built-in module designed to facilitate the processing of Python abstract syntax grammar trees. Our aim is to retain the core components of the code, specifically class and function modules. However, if the token size of the executed core code still exceeds our imposed limitations, we turn to the latest GPT models to generate code summaries.

In a study conducted by Sun and Chen \cite{sun:chen}, various code summarization models, including ChatGPT, were evaluated. ChatGPT, which is fine-tuned from a GPT-3.5 series model through reinforcement learning based on human feedback, demonstrates noteworthy strengths. As highlighted by \cite{sun:chen}, ChatGPT exhibites an enhanced ability to grasp the underlying meaning of the code, unaffected by the code's specific vocabulary. Additionally, ChatGPT tends to provide more detailed descriptions of code behavior in the generated comments.

In our project, we leverage the capabilities of the cutting-edge GPT model, specifically the \emph{gpt-4-32k} model, for code summarization. The \emph{gpt-4-32k} model accommodates input sizes of up to 32,000 tokens and excels at understanding code semantics. We use a simple prompt to generate summaries: ``Generate a summary for the following code: \texttt{\textbackslash n} $<code>$".

\paragraph{Data Storage and Queries}

After finalizing the embedding generation process, data objects and their corresponding embeddings are inserted into a vector database named Weaviate (see Table \ref{schema:weaviate}). To ensure data consistency with the Jupyter Notebook Repository, the database will be updated every 15 minutes. Subsequently, our search engine accommodates various query types, including Exact Query (EQ), User Defined Query (UDQ), and Code Summary Query (CSQ).

The first query type, EQ, is used to test whether the original cell can be retrieved by inputting partial code or text copied and pasted from notebooks. The second query type, UDQ, offers greater flexibility, allowing users to express their requests and retrieve relevant content accordingly. Lastly, CSQ, a form of `reverse engineering',  is employed to assess the system's ability to retrieve original code blocks with code summaries (in particular, to preserve model consistency, we use the same \emph{gpt-4-32k} model to generate the code summary).

These three query types represent distinct perspectives for understanding notebooks, and we postulate that the search process for each query type serves as a robustness test for large language models, guarantee their ability to comprehend semantics across varying levels of complexity.

Due to data privacy considerations, we avoid disclosing specific contents utilized in semantic search. Instead, we present a table depicting the distances between query results generated by different query inputs and the relevant notebooks (see Table \ref{semantic:distance}). We figure out that the distance (calculated by 1-cosine similarity) between the code summaries generated by the GPT model and the actual content is remarkably close. This astonishing finding suggests that GPT models might possess a deeper understanding of content semantics compared to human searchers and even the original authors themselves!

\begin{table}
\small
\caption{Schema Design}\label{schema:weaviate}
\begin{tabular}{|p{1.7cm}|p{1.3cm}|p{1.2cm}|p{2cm}|p{1.8cm}|p{1.6cm}|p{1.3cm}|}
\hline
notebook\_id & contents & cell\_type & author\_name & modified\_at & created\_at & vector\\
\hline
``string" & ``text" & ``string" & ``string" & ``number" & ``number" & ``list of floats"\\
\hline
Notebook path & cell contents & [text, code] & author name & modification time & creation time & OpenAI Client Embedding\\
\hline
\end{tabular}
\end{table}

\begin{table}
\centering
\caption{Distances between Queries and Retrieved Contents}\label{semantic:distance}
\begin{tabular}{|l|l|l|}
\hline
Query Types &  Perspective of Understanding & Distance\\
\hline
Exact Query &  Author & 0.116\\
User Defined Query & Searcher & 0.163\\
Code Summary Query & GPT & \textbf{0.065}\\
\hline  
\end{tabular}
\end{table}

\section{Conclusions and Future Work}

In this study, we have delved into the potential of applying large language models (LLMs) to semantic searches within Jupyter Notebooks. Our findings confirm that these models exhibit a remarkable capability to comprehend both textual and code-based data. They are adept at capturing the intricate internal mechanisms and profound logic within the code, summarizing semantics in a well-structured manner.

For future work, we have identified several key objectives: 

\begin{itemize}
    \item Augmenting data inputs for LLMs: We plan to enrich the information provided to LLMs by including additional details about cell contents, such as file paths, cell types, and author names. 
    \item In addition to generating embeddings for Markdown-type and Code-type cells, one potential task is to include outputs and graphs, which require a more advanced multimodal LLMs.
    \item Exploring code summarization at the cell level: We intend to investigate the impact of applying code summarization to all cells without decomposing them into function-level segments, aiming to assess its effect on search results. 
    \item Conducting qualitative and quantitative analyses: We will develop a comprehensive set of both qualitative and quantitative analyses based on the search results. These analyses will scrutinize how robust and adaptable embeddings generated by LLMs are when dealing with codes of varying 'quality'.
\end{itemize}

All these analyses will contribute to understand and evaluate how robust and adaptable LLM-generated embeddings are when dealing with codes of varying ``quality" and across different programming languages.



\begin{thebibliography}{9}
%
\bibitem{muen}
Muennighoff, N. (2022). Sgpt: Gpt sentence embeddings for semantic search. arXiv preprint arXiv:2202.08904.

\bibitem{nayak}
Nayak, P. (2019). Understanding searches better than ever before.

\bibitem{jeff:cass}
Jeffrey Zhu, Mingqin Li, Jason Li, and Cassandra Oduola. (2021). Bing delivers more contextualized search using quantized transformer inference on NVIDIA GPUs in Azure.

\bibitem{koubaa:latif}
Koubaa, A., Boulila, W., Ghouti, L., Alzahem, A., \& Latif, S. (2023). Exploring ChatGPT capabilities and limitations: A critical review of the nlp game changer.

\bibitem{liu:ge}
Liu, Y., Han, T., Ma, S., Zhang, J., Yang, Y., Tian, J., ... \& Ge, B. (2023). Summary of ChatGPT-Related Research and Perspective Towards the Future of Large Language Models. Meta-Radiology, 100017.

\bibitem{gu:gall}
Gu, J., Salza, P., \& Gall, H. C. (2022, March). Assemble foundation models for automatic code summarization. In 2022 IEEE International Conference on Software Analysis, Evolution and Reengineering (SANER) (pp. 935-946). IEEE.

\bibitem{klu:will}
Kluyver, T., Ragan-Kelley, B., Pérez, F., Granger, B. E., Bussonnier, M., Frederic, J., ... \& Willing, C. (2016). Jupyter Notebooks-a publishing format for reproducible computational workflows. Elpub, 2016, 87-90.


\bibitem{weaviate}
Weaviate: an open-source vector database. (2023). Retrieved from https://weaviate.io/.

\bibitem{sun:chen}
Sun, W., Fang, C., You, Y., Miao, Y., Liu, Y., Li, Y., ... \& Chen, Z. (2023). Automatic Code Summarization via ChatGPT: How Far Are We?. arXiv preprint arXiv:2305.12865.

\bibitem{fl:bl}
Fang, L., Li, L., Liu, Y., Torvik, V. I., and Ludäscher, B. (2023). Kaer: A knowledge
augmented pre-trained language model for entity resolution. Knowledge Augmented
Methods for Natural Language Processing workshop in conjunction with AAAI 2023.

\bibitem{morales}
Morales, C., Collarana, D., Vidal, M. E., \& Auer, S. (2017). Matetee: A semantic similarity metric based on translation embeddings for knowledge graphs. In Web Engineering: 17th International Conference, ICWE 2017, Rome, Italy, June 5-8, 2017, Proceedings 17 (pp. 246-263). Springer International Publishing.

\bibitem{meyer}
Meyer, L. P., Stadler, C., Frey, J., Radtke, N., Junghanns, K., Meissner, R., ... \& Martin, M. (2023). Llm-assisted knowledge graph engineering: Experiments with chatgpt. arXiv preprint arXiv:2307.06917.

\end{thebibliography}
\end{document}